\documentclass{article}

\usepackage{amssymb}
\usepackage{amsfonts}
\usepackage{epsfig}

\newcommand{\Au}{{\bf (A.1)}}
\newcommand{\Ad}{{\bf (A.2)}}
\newcommand{\At}{{\bf (A.3)}}
\newcommand{\Aq}{{\bf (A.4)}}
\newcommand{\Ac}{{\bf (A.5)}}

\newcommand{\toP}{\stackrel{P}{\to}}
\newcommand{\tod}{\stackrel{d}{\to}}

\newcommand{\toPC}{\stackrel{c}{\longrightarrow}}

\newcommand{\R}{\mathbb{R}}
\newcommand{\PP}{\mathbb{P}}
\newcommand{\E}{\mathbb{E}}
\newcommand{\V}{\mathbb{V}}

\newcommand{\CQFD}
{%
\mbox{}%
\nolinebreak%
\hfill%
\rule{2mm}{2mm}%
\medbreak%
\par%
}
 \newcommand{\proof}{{\bf Proof~: }}

\newtheorem{Theo}{Theorem}

\newtheorem{Coro}{Corollary}
\newtheorem{Lem}{Lemma}
\newtheorem{Remark}{Remark}

\begin{document}

\title{Frontier estimation via kernel regression on high power-transformed data}
\author{S\'ephane Girard$^{(1)}$ \& Pierre Jacob$^{(2)}$}
\date{
$^{(1)}$INRIA Rh\^one-Alpes, team Mistis, Inovall\'ee, 655, av. de l'Europe, Montbonnot, 38334 Saint-Ismier cedex, France, {\tt Stephane.Girard@inrialpes.fr}
\\
$^{(2)}$Universit\'e Montpellier 2, EPS-I3M, place Eug\`ene Bataillon,\\
 34095 Montpellier cedex 5, France, {\tt jacob@math.univ-montp2.fr}\\
}
\maketitle
\begin{abstract}
We present a new method for estimating the frontier of a multidimensional
sample.  
The estimator is based on a kernel regression on the
power-transformed data. We assume that the exponent of the transformation
goes to infinity while the bandwidth of the kernel goes to zero.  
We give conditions on these two parameters to obtain
complete convergence and asymptotic normality.
The good performance of the estimator is
illustrated on some finite sample situations.


\end{abstract}

\section{Introduction}
%--------------------------------------------------------------
\label{intro}

Let $(X_i,Y_i)$, $i=1,\dots,n$ be independent
copies of a random pair $(X,Y)$ with support $S$ defined by
\begin{equation}
\label{defS}
S=\{(x,y)\in E\times\R; 0\leq y \leq g(x)\}.
\end{equation}
The unknown function $g:E\to\R$ is called the frontier.
We address the problem of estimating $g$ in the case $E=\R^d$.  
Our estimator of the frontier is based on a kernel regression on
the power-transformed data. More precisely, the estimator of  
$g$ is defined for all $x\in\R^d$ by
\begin{equation}
\label{defest}
\hat g_n(x)=\left((p+1) \left.\sum_{i=1}^n K_h(x-X_i) Y_i^p \right/ 
\sum_{i=1}^n K_h(x-X_i) \right)^{1/p},
\end{equation}
where $p=p_n$ and $h=h_n$ are non random sequences such that 
$h\to 0$ and $p\to\infty$ as $n\to\infty$. 
This latter condition is the key so that the high power-transformed
data ``concentrate'' along the frontier.  
We have also introduced  $K_h(t)=K(t/h)/h^d$
where $K$ is a probability density function (pdf) on $\R^d$.
In this context, $h$ is called the window-width.

From the practical point of view, note that, compared to the
extreme value based 
estimators~\cite{ISUPLaurent,Geffroy,Scandi,JSPI,JSPI2,ESAIM}, 
projection estimators~\cite{JacSuq}
or piecewise polynomial estimators~\cite{KorTsy,KorTsy3,Har},
this estimator does not require a partition of $S$
and is thus not limited to bi-dimensional bounded supports.  
Moreover, it benefits from an explicit formulation which is not
the case of estimators defined by optimization problems~\cite{Russe2}
such as local polynomial estimators~\cite{Hall,Hall3,Keith}.
From the theoretical point of view, this estimator reveals to be
completely convergent to $g$ without assumption neither on the
distribution of $X$ nor on the distribution of $Y$ given $X=x$
(see Section~\ref{acc}).
Note however that $(p+1)^{1/p}\to 1$ when $p\to\infty$.
In fact, this correcting term is specially designed  for the case
where $Y$ given $X=x$ is uniformly distributed on $[0,g(x)]$.
In this latter situation,
the estimator is asymptotically Gaussian with the 
rate of convergence $n^{-\alpha/(d+\alpha)}$
(see Section~\ref{an}).
This rate is proved to be minimax optimal 
for $\alpha-$ Lipschitzian $d-$ dimensional frontiers~\cite{KorTsy}, Chapter~5.
This result is generalized in~\cite{MamTsy} to 
boundaries of more general regions.  Other extensions
are provided in~\cite{Hall2, Har} to densities of $Y$ given $X=x$
decreasing as a power of the distance from the boundary.  
We refer to~\cite{DST,Farrel,Gijbels2} for the estimation of frontier functions under monotonicity assumptions, and to~\cite{Aragon,Cazals} for the definition
of robust estimators in this context.
We conclude this paper by an illustration of the behavior
of our estimator on some finite sample situations in Section~\ref{simul}
and by describing our future work in Section~\ref{futur}.
Technical lemmas are postponed to the appendix.  

\section{Notations and assumptions}
%--------------------------------------------------------------

To motivate the estimator~(\ref{defest}), consider the random
variable $Z=(p+1)Y^p$ and the conditional expectation
$r_n(x)=\E(Z|X=x)$. 
Estimating the frontier $g$ is often related to estimating the regression function $r_n$.
For instance, if $Y$ given $X=x$ is uniformly distributed
on $[0,g(x)]$, we have $r_n^{1/p}(x)=g(x)$. 
A similar remark is done in~\cite{JacSuq2} where regression estimators
are modified to build estimators of the frontier,
but the profound difference here is that $p\to\infty$.
This condition allows to obtain $r_n^{1/p}(x)\to g(x)$
even when $Y$ given $X=x$ is not uniformly distributed 
(see Lemma~\ref{sauvetage} below).
We denote by $f$ the pdf of the random
vector $X$ and we introduce 
\begin{equation}
\label{defest1}
\hat \varphi_n(x)= \frac{1}{n} \sum_{i=1}^n K_h(x-X_i) Z_i,
\end{equation}
where $Z_i=(p+1)Y_i^p$. Note that $\hat \varphi_n(x)$
can be seen as a classical kernel estimator of 
$\varphi_n(x)=f(x)r_n(x)$ but 
keep in mind that $p\to\infty$. Similarly,
\begin{equation}
\label{defest2}
\hat f_n(x)= \frac{1}{n} \sum_{i=1}^n K_h(x-X_i)
\end{equation}
is an estimator of $f(x)$ and 
\begin{equation}
\label{defest3}
\hat r_n(x)= \hat \varphi_n(x) / \hat f_n(x)
\end{equation}
is an estimator of $r_n(x)$.
Collecting (\ref{defest1}), (\ref{defest2}) and (\ref{defest3}),
our estimator (\ref{defest}) can be rewritten 
as $$\hat g_n(x)=\hat r_n(x)^{1/p}.$$
To establish the asymptotic properties of $\hat g_n(x)$, the
following assumptions are considered:

\Au:  $g$ is $\alpha$-Lipschitz, $f$ is $\beta$-Lipschitz, with 
$0<\alpha\leq\beta\leq 1$,

\Ad: $0< g_{\min}\leq g(x)$, $\forall x\in\R^d$,

\At: $f(x)\leq f_{\max}<\infty$, $\forall x\in\R^d$,

\Aq:  $K$ is a Lipschitzian pdf on $\R^d$, with support included in $B$,
the unit ball of $\R^d$.

\noindent Note that \Aq~implies that, for all $q\geq 1$, we have
$0<\int_B K^q(x)dx < +\infty$.

\section{Complete convergence}
%--------------------------------------------------------------
\label{acc}

In this section, the complete convergence of the frontier estimator
toward the true frontier is established. The next lemma can be seen as
the intuitive justification why no assumption on the conditional distribution of $Y$ given $X$ is required in the proof of Theorem~\ref{ACC}.
\begin{Lem}
\label{sauvetage}
Under \Ad, for all $x\in B$, $r_n(x)^{1/p}\to g(x)$ as $n\to\infty$.
\end{Lem}
\proof Let $\varepsilon>0$. Since $(X,Y)$ has support $S$ defined
 by~(\ref{defS}), it follows that
$$
r_n(x)=(p+1)\E\left(Y^p|X=x\right)\leq (p+1) g^p(x)
$$
and thus, since $(p+1)^{1/p}\to 1$ as $p\to\infty$, for $n$ large
enough and all $x\in B$,
\begin{equation}
\label{eqp1}
r_n^{1/p}(x)\leq (1+\varepsilon) g(x).
\end{equation}
Moreover, we have,
\begin{eqnarray*}
r_n(x)&\geq& (p+1)\E\left(Y^p {\mathbf 1}\{ Y> g(x)-\varepsilon \} |X=x\right)\\
&\geq & (p+1) (	g(x)-\varepsilon )^p \PP( Y> g(x)-\varepsilon |X=x).
\end{eqnarray*}
Now, since  $(X,Y)$ has support $S$, one can assume without loss of
generality that $Y$ given $X=x$ has support $[0,g(x)]$ such that
$ \PP( Y> g(x)-\varepsilon |X=x)>0$. It follows that 
$$
\left[(p+1)  \PP( Y> g(x)-\varepsilon |X=x) \right]^{1/p}\to 1
$$
as $p\to\infty$, and consequently, for $n$ large enough,
\begin{equation}
\label{eqp2}
r_n^{1/p}(x)\geq (1-\varepsilon) g(x).
\end{equation}
Collecting~(\ref{eqp1}) and~(\ref{eqp2}) gives the result.
\CQFD
\begin{Theo}
\label{ACC}
Suppose \Au--\Aq~hold and $nh^d/\log n\to\infty$. Then $\widehat{g}_{n}(x)$
converges completely to $g(x)$ for all $x\in\R^d$
such that $f(x)>0$.  
\end{Theo}
\proof Let $x\in\R^d$ such that $f(x)>0$ and let $\varepsilon$ such that
$0<\varepsilon<g(x)$.  Define $0<\eta<1/4$ by $\eta=\varepsilon/(4g(x))$. Then,
from Lemma~\ref{lemcinq},
\begin{eqnarray*}
\left\{  \left\vert \widehat{g}_{n}(x)-g(x)\right\vert >\varepsilon\right\}
&=&\left\{  \left\vert  \frac{\widehat{r}_{n}^{1/p}(x)}{g(x)}
-1\right\vert >4\eta\right\} \\ 
&\subseteq&
\left\{  \left\vert \left(  \frac{\widehat{\varphi}_{n}(x)}{f(x)g^p(x)}\right)
^{1/p}-1\right\vert >\eta\right\} 
\cup
\left\{  \left\vert \left(  \frac{\hat{f}_{n}(x)}{f(x)}\right)^{1/p}
 -1\right\vert >\eta\right\}. 
\end{eqnarray*}
Since $\hat f_n(x)$ converges completely to $f(x)$, see e.g.~\cite{Bosq},
Chapter~4, Theorem~III.3, it follows that $(\hat f_n(x)/f(x))^{1/p}$ converges
completely to 1.
Therefore, writing
$$
\left(\frac{\widehat \varphi_n(x)}{f(x)g^p(x)}\right)^{1/p}=(p+1)^{1/p}T_n(x)$$
with
$$
T_{n}(x)=\left[  \frac{1}{n}\sum_{i=1}^{n}K_{h}(x-X_{i})\left(
\frac{Y_{i}}{g(x)}\right)^{p}\frac{1}{f(x)}\right]  ^{1/p}
$$
and remarking that $(p+1)^{1/p}\to 1$ as $n\to\infty$, it suffices to consider 
$$
\left\{  \left\vert T_{n}(x)-1\right\vert >\eta\right\}  \subseteq\left\{
T_{n}(x)>1+\eta\right\}  \cup\left\{  T_{n}(x)<1-\eta\right\}.  
$$
The two events are studied separately.
First, let $0<\delta<\eta$. Then, $\|x-X_i\|\leq h$ entails
$$
Y_i-g(x)(1+\delta)  \leq  g(X_i)-g(x) -\delta g(x)  \leq L_g h^\alpha - \delta
g_{\min} < 0
$$
for $n$ large enough and where $L_g$ is the Lipschitz constant
associated to $g$. We thus have
\begin{eqnarray*}
T_{n}(x)&=&\left[  \frac{1}{n}\sum_{i=1}^{n}K_{h}(x-X_{i})\left(
\frac{Y_{i}}{g(x)}\right)^{p}\mathbf{1}\{Y_i<g(x)(1+\delta)\} \frac{1}{f(x)}\right]  ^{1/p}\\
&\leq& 
(1+\delta) \left[
\frac{1}{n}\sum_{i=1}^{n}K_{h}(x-X_{i}) \mathbf{1}\{Y_i<g(x)(1+\delta)\} \frac{1}{f(x)}\right]  ^{1/p},
\end{eqnarray*}
and consequently,
\begin{eqnarray*}
\{ T_{n}(x)>1+\eta\} &\subseteq& \left\{ 
\frac{1}{n}\sum_{i=1}^{n}K_{h}(x-X_{i}) \mathbf{1}\{Y_i<g(x)(1+\delta)\} \frac{1}{f(x)}
> \left(\frac{1+\eta}{1+\delta}\right)^p
\right\}\\
 &\subseteq& \left\{ 
\frac{1}{n}\sum_{i=1}^{n}K_{h}(x-X_{i}) \mathbf{1}\{Y_i<g(x)(1+\delta)\} \frac{1}{f(x)}
> 2
\right\},
\end{eqnarray*}
since, for $n$ large enough, $((1+\eta)/(1+\delta))^p>2$.
From~\cite{Bosq}, Chapter~5, Corollary~II.4, the following complete
convergence holds:
$$
\frac{1}{n}\sum_{i=1}^{n}K_{h}(x-X_{i}) \mathbf{1}\{Y_i<g(x)(1+\delta)\} \frac{1}{f(x)}
\toPC \PP(Y < g(x)(1+\delta) | X=x) =1,
$$
and therefore
$$
\sum_{n=1}^\infty \PP(T_n(x)>1+\eta) \leq
\sum_{n=1}^\infty \PP\left(\frac{1}{n}\sum_{i=1}^{n}K_{h}(x-X_{i}) \mathbf{1}\{Y_i<g(x)(1+\delta)\} \frac{1}{f(x)}-1 >1
\right)< +\infty,
$$
which concludes the first part of the proof.
Second,
\begin{eqnarray*}
T_{n}(x)&\geq&\left[  \frac{1}{n}\sum_{i=1}^{n}K_{h}(x-X_{i})\left(
\frac{Y_{i}}{g(x)}\right)^{p}\mathbf{1}\{Y_i>g(x)(1-\delta)\} \frac{1}{f(x)}\right]  ^{1/p}\\
&\geq& 
(1-\delta) \left[
\frac{1}{n}\sum_{i=1}^{n}K_{h}(x-X_{i}) \mathbf{1}\{Y_i>g(x)(1-\delta)\} \frac{1}{f(x)}\right]  ^{1/p},
\end{eqnarray*}
and consequently,
$$
\{ T_{n}(x)<1-\eta\} \subseteq \left\{ 
\frac{1}{n}\sum_{i=1}^{n}K_{h}(x-X_{i}) \mathbf{1}\{Y_i>g(x)(1-\delta)\} \frac{1}{f(x)}
< \left(\frac{1-\eta}{1-\delta}\right)^p
\right\}.
$$
Now, since $\PP(Y>g(x)(1-\delta)|X=x)>0$, there exists $\gamma>0$ such that,
for $n$ large enough, 
$$
\left(\frac{1-\eta}{1-\delta}\right)^p - \PP(Y>g(x)(1-\delta)|X=x) < -\gamma,
$$
entailing that, for $n$ large enough,
$$
\{ T_{n}(x)<1-\eta\} \subseteq \left\{ 
\frac{1}{n}\sum_{i=1}^{n}K_{h}(x-X_{i}) \mathbf{1}\{Y_i>g(x)(1-\delta)\} \frac{1}{f(x)}
- \PP(Y/g(x)>1-\delta|X=x) < -\gamma
\right\}.
$$
Taking into account of the following complete convergence 
$$
\frac{1}{n}\sum_{i=1}^{n}K_{h}(x-X_{i}) \mathbf{1}\{Y_i>g(x)(1-\delta)\} \frac{1}{f(x)}
\toPC \PP(Y > g(x)(1-\delta) | X=x),
$$
it follows that
\begin{eqnarray*}
&&\sum_{n=1}^\infty \PP(T_n(x)<1-\eta)\\ 
&\leq&
\sum_{n=1}^\infty \PP\left(
\frac{1}{n}\sum_{i=1}^{n}K_{h}(x-X_{i}) \mathbf{1}\{Y_i>g(x)(1-\delta)\} \frac{1}{f(x)}
- \PP(Y>g(x)(1-\delta)|X=x) < -\gamma
\right)\\
&<& +\infty,
\end{eqnarray*}
which concludes the second part of the proof.
\CQFD

\section{Asymptotic normality}
%--------------------------------------------------------------
\label{an}

\noindent Second, the asymptotic normality of the frontier estimator
centered on the true frontier is established.  
To this end, asymptotic expansions
of the expectation and variance of $\hat\varphi_n(x)$ are needed.  
These calculations are done under the additional assumption 

\Ac: $Y$ given $X=x$ is uniformly distributed on $[0,g(x)]$.

\noindent The next two lemmas are similar to classical ones in kernel
regression (see for instance~\cite{LivreToulouse}, Theorem~6.11),
but the dependence on $n$ of the function $\varphi_n(x)$
induces technical difficulties.  
We first establish that $\widehat{\varphi}_{n}(x)$ is an asymptotically
unbiased estimator of $\varphi_{n}(x)$ in the sense that 
$\E\widehat{\varphi}_{n}(x)/\varphi_{n}(x)\to 1$ as $n\to\infty$
provided that $ph^\alpha\to0$.

\begin{Lem}
\label{lemun}
Under \Au--\Ac, if $ph^\alpha\to0$, then for all $x\in\R^d$
$$
\E\widehat{\varphi}_{n}(x)=\varphi_{n}(x)\left[  1+O(ph^{\alpha})\right].
$$
\end{Lem}

\noindent\proof From (\ref{defest1}), it follows that
$$
\E\widehat{\varphi}_{n}(x)=\E(K_{h}(x-X)Z)=\E(K_{h}(x-X)\E(Z|X)),
$$
so that, by a straightforward calculation, and recalling that
$\varphi_{n}(u)=g^{p}(u)f(u)$, we obtain
\begin{eqnarray}
\label{similar}
\E\widehat{\varphi}_{n}(x)&=&\E(K_{h}(x-X)g^{p}(X))=\int_{\R^d} \frac{1}{h^d}K\left(\frac{x-u}{h}\right)\varphi_{n}(u)du \\
&=& \int_B K(y)  \varphi_{n}(x-hy) dy, \nonumber
\end{eqnarray}
with a classical change of variable, and since $K$ has a  
compact support. We thus can write:
$$
\E\widehat{\varphi}_{n}(x)-\varphi_{n}(x)=\int_B K(y)\left[  \varphi
_{n}(x-hy)-\varphi_{n}(x)\right]  dy.
$$
Consider now the decomposition below:
$$
|\varphi_{n}(x-hy)-\varphi_{n}(x)|\leq f(x-hy)\left|  g^{p}(x-hy)-g^{p}(x)\right|
+g^{p}(x)\left|  f(x-hy)-f(x)\right| := T_1 + T_2. 
$$
Following Lemma~\ref{lemquatre}, 
\begin{eqnarray*}
T_1&=&f(x-hy)g^{p}(x)\left|  \frac{g^{p}(x-hy)}{g^{p}(x)}-1\right|
\leq2f_{\max}\frac{L_{g}}{g_{\min}}g^{p}(x)ph^{\alpha}=g^{p}(x)O(ph^{\alpha}),\\
T_2&\leq& g^{p}(x)L_{f}h^{\beta}=g^{p}(x)O(h^{\beta})=g^p(x)o(ph^{\alpha}),
\end{eqnarray*}
where $L_{f}$ and $L_{g}$ are the Lipschitz constants of the functions $f$ and
$g$. Finally,
$$
\E\widehat{\varphi}_{n}(x)-\varphi_{n}(x)=g^{p}(x)O(ph^{\alpha})=\varphi
_{n}(x)O(ph^{\alpha}),
$$
and the conclusion follows.  
\CQFD
\noindent Similarly, we now provide an equivalent expression for 
$\V(\widehat{\varphi}_{n}(x)/{\varphi}_{n}(x))$
which appears to be of order $p/(nh^d)$.
Thus, condition  $p/(nh^d)\to 0$ will be necessary in Corollary~\ref{coro}
to obtain the weak consistency of $\widehat{\varphi}_{n}(x)$,
i.e. to ensure that 
$\widehat{\varphi}_{n}(x)/{\varphi}_{n}(x)\toP 1$.

\begin{Lem}
\label{lemdeux}
Under \Au--\Ac, if $ph^\alpha\to0$ then for all $x\in\R^d$,
$$
\V(\widehat{\varphi}_{n}(x))=\frac{1}{nh^d}\frac{(p+1)^{2}}{2p+1}\int_B
K^{2}(s)ds\frac{\varphi_{n}^{2}(x)}{f(x)}\left[  1+o(1)\right].
 $$
\end{Lem}
\proof We have
\begin{eqnarray*}
\V(\widehat{\varphi}_{n}(x))&=&\frac{1}{n^{2}}\sum_{i=1}^{n}\V\left(
K_{h}(x-X_{i})Z_{i}\right)  =\frac{1}{n}\V\left(  K_{h}(x-X)Z\right) \\ 
&=&
\frac{1}{nh^{2d}}\E\left(  K^{2}\left(\frac{x-X}{h}\right)Z^{2}\right)
-\frac{1}{n} \E^{2}\left(  \widehat{\varphi}_{n}(x)\right) :=T_{3}+T_{4}.
\end{eqnarray*}
From Lemma~\ref{lemun}, we immediately derive 
$$
T_{4}=\frac{1}{n}\varphi_n^{2}(x)\left[  1+o(1)\right].
$$ 
We shall prove that  
\begin{equation}
\label{eqT3}
T_{3}=\frac{1}{nh^d} \frac{(p+1)^{2}}{2p+1}\int_B K^{2}(s)ds\frac{\varphi_{n}^{2}(x)}{f(x)}\left[ 1+o(1)\right], 
\end{equation}
leading to $T_{4}/T_{3}=O({h}/{p})$,  and the announced result follows.
To this end, remark that
\begin{eqnarray*}
T_{3}&=&\frac{1}{nh^{2d}}\E\left(  K^{2}\left(\frac{x-X}{h}\right)\E(Z^{2}|X)\right)\\
&=& \frac{1}{nh^{2d}}\frac{(p+1)^{2}}{2p+1}\E\left(  K^{2}\left(\frac{x-X}
{h}\right)g^{2p}(X)\right)  \\
&=&\frac{1}{nh^d}\frac{(p+1)^{2}}{2p+1}\int_B K^{2}(s)ds\int_{\R^d}\frac{1}{h^d}
Q\left(\frac{x-u}{h}\right)g^{2p}(u)f(u)du,
\end{eqnarray*}
where we have introduced the kernel $Q=K^{2}/\int_B K^{2} (s)ds$.
It is easily seen that the second integral is similar to this
appearing in $\E\widehat{\varphi}_{n}(x)$, (see (\ref{similar})), 
with $K$ replaced by $Q$ and $p$ by $2p$.
Thus, as in the proof of Lemma~\ref{lemun}, we have
$$
\int_{\R^d}\frac{1}{h^d}Q\left(\frac{x-u}{h}\right)g^{2p}(u)f(u)du=g^{2p}(x)f(x)\left[
1+o(1)\right]  =\frac{\varphi_{n}^{2}(x)}{f(x)}\left[  1+o(1)\right] ,
$$
and (\ref{eqT3}) is proved.  
\CQFD
\noindent As a simple consequence of Lemma~\ref{lemun} and Lemma~\ref{lemdeux}, we have
\begin{Coro}
\label{coro}
Under \Au--\Ac, if $ph^\alpha\to0$ and $p/(nh^d)\to0$, then, for all $x\in\R^d$,
$$\widehat\varphi_n(x)/\varphi_n(x)\toP 1.$$
\end{Coro}
We can now turn to our main result.  
\begin{Theo}
\label{AN}
Suppose that $nph^{d+2\alpha}\rightarrow0$ and $p/(nh^d)\rightarrow0$.
Let us define
$$
\sigma_{n}^{-1}(x)=((2p+1)nh^d)^{1/2}
\left(\frac{f(x)}{\int_B K^{2}(t)dt}\right)^{1/2}.
$$
Then, under \Au--\Ac, for all $x\in\R^d$,
$$
\sigma_{n}^{-1}(x)\left(  \frac{\widehat{g}_{n}(x)}{g(x)}-1\right) \tod
N(0,1).
$$
\end{Theo}
\proof
First, note that $nph^{d+2\alpha}\rightarrow0$ and $p/(nh^d)\rightarrow0$
imply $ph^\alpha\to 0$.
From Lemma~\ref{lemhuit}, it suffices to prove that $$
\xi_{n}:=\frac{\sigma_{n}^{-1}(x)}{p}\left(  \frac{\widehat{\varphi}_{n}(x)}{\varphi
_{n}(x)}-\frac{\E\widehat{\varphi}_{n}(x)}{\varphi_{n}(x)}\right)  
\tod N(0,1).
$$
To this end, define 
$$
W_{i,n}=\frac{\sigma_{n}^{-1}(x)}{np}\frac{1}{\varphi_{n}(x)}
K_{h}(x-X_{i})Z_{i}
$$
so that  we can write
 $$
\xi_{n}=\sum_{i=1}^{n}\left(W_{i,n}-\E W_{i,n}\right).
$$
Following Lemma~\ref{lemdeux}, we have
\begin{eqnarray*}
\V\left(  \xi_{n}\right) &=&n\V\left(  W_{1,n}\right)  =\frac{\sigma
_{n}^{-2}(x)}{p^{2}}\frac{1}{\varphi_{n}^{2}(x)}\V(\widehat{\varphi}_{n}(x))
 \\
&=&\frac{\left(  2p+1\right)  nh^d f(x)}{p^2\int_B K^{2}(s)ds }\frac{1}{\varphi_{n}
^{2}(x)}\frac{1}{nh^d}\frac{(p+1)^{2}}{2p+1}\int_B K^{2}(s)ds \frac{\varphi_{n}^{2}
(x)}{f(x)}\left[  1+o(1)\right] 
 \\
 &=&  1+o(1). 
\end{eqnarray*}
Thus, the condition of Lyapounov reduces to
\begin{equation}
\sum_{i=1}^{n}\E\left\vert W_{i,n}-\E W_{i,n}\right\vert ^{3}
=n\E\left\vert W_{1,n}-\E W_{1,n}\right\vert ^{3}\rightarrow0.
\label{etoile2}
\end{equation}
Taking into account that $W_{1,n}$ is a positive random variable,
the triangular inequality together with Jensen's inequality yield
$$
\E\left\vert W_{1,n}-\E W_{1,n}\right\vert ^{3} \leq 
8 \E\left(W_{1,n}^3\right).
$$
Introducing the kernel $K^{3}/\int_B K^{3}(s)ds$, and mimicking
the proof of Lemma~\ref{lemdeux}, we obtain
\begin{equation}
\label{mom3}
\E(W_{1,n}^{3})= \frac{n^{-3/2}h^{-d/2}p^{1/2}2^{3/2}}{3{f(x)}^{3/2}}
\frac{\int_B K^{3}(s)ds }{\left(  \int_B K^{2}(s)ds\right)  ^{3/2}}(1+o(1))
 =\kappa n^{-3/2}h^{-d/2}p^{1/2}(1+o(1)),
\end{equation}
where $\kappa$ is a positive constant.  Returning to (\ref{etoile2}), we have
$$
\sum_{i=1}^{n}E\left\vert W_{i,n}-\E W_{i,n}\right\vert ^{3}
\leq 8 \kappa\left(\frac{p}{nh^d}\right)^{1/2}(1+o(1))
\rightarrow0,
$$
and the result is proved.
\CQFD
\begin{Remark}
Theorem~\ref{AN} holds when $\sigma_n^{-1}(x)$ is replaced with
 $$
\widehat\sigma_{n}^{-1}(x)=((2p+1)nh^d)^{1/2}
\left(\frac{\widehat f_n(x)}{\int_B K^{2}(t)dt}\right)^{1/2},
$$
since in this context $\widehat f_n(x)\toP f(x)$.  This allows
to produce pointwise confident intervals for the frontier.  
\end{Remark}
\begin{Remark}
\label{rqun}
To fulfill the assumptions of Theorem~\ref{AN}, one can choose
$h=n^{-1/(d+\alpha)}$ and $p=\varepsilon_n n^{\alpha/(d+\alpha)} $,
where $(\varepsilon_n)$ is a sequence tending to zero arbitrarily
slowly.  These choices yield 
$$
\sigma_n^{-1}(x)= \varepsilon_n^{1/2} n^{\alpha/(d+\alpha)}
\left(\frac{2f(x)}{\int_B K^{2}(t)dt}\right)^{1/2}(1+o(1)),
$$
 which is the optimal speed (up to the
 $ \varepsilon_n$ factor) for estimating $\alpha-$ Lipschitzian
 $d-$ dimensional frontiers, see~\cite{KorTsy}, Chapter~5.  
\end{Remark}
The good performances of $\hat g_n(x)$ on finite sample situations
are illustrated in the next section. Remark~\ref{rqun} will be of
great help to choose $p$ and $h$ sequences.  

%--------------------------------------------------------------------
\section{Numerical experiments}
%--------------------------------------------------------------------
\label{simul}

Here, we limit ourselves to unidimensional random variables $X$
($p=1$) with compact support $E=[0,1]$. Besides, $Y$ given $X=x$
is distributed on $[0,g(x)]$ such that
\begin{equation}
\label{proba}
\PP(Y>y|X=x)=\left(1-\frac{y}{g(x)}\right)^\gamma,
\end{equation}
with $\gamma>0$. This conditional survival distribution function
belongs to the Weibull domain of attraction, with extreme value
index $-\gamma$, see~\cite{EMBR} for a review on this topic.
The case $\gamma=1$ corresponds to the situation where $Y$ given $X=x$
is uniformly distributed on $[0,g(x)]$. The larger $\gamma$ is,
the smaller the probability~(\ref{proba}) is, when $y$ is close to
the frontier $g(x)$.
The behavior of the proposed frontier estimator is investigated
on different situations:
\begin{itemize}
\item Two distributions are considered for $X$: a uniform distribution
$U([0,1])$ and a beta distribution $B(2,2)$.
\item Two frontiers are introduced. The first one
$$
\begin{array}{l}
g_1(x)=\\
\\
\\
\end{array}
\left|
\begin{array}{ll}
1+\exp{(-60(x-1/4)^2)}& \mbox{ if } 0\leq x \leq 1/3, \\
1+\exp{(-5/12)} & \mbox{ if } 1/3 < x \leq 2/3, \\
1+5\exp{(-5/12)}-6\exp{(-5/12)}x & \mbox{ if } 2/3 < x \leq 5/6, \\
6x-4 & \mbox{ if } 5/6 < x \leq 1.
\end{array}
\right.
$$
is continuous but is not derivable at
$x=1/3$, $x=2/3$ and $x=5/6$.
The second one
$$
g_2(x)=(1/10+\sin(\pi x))\left(11/10-\exp\left(-64(x-1/2)^2\right)/2\right)
$$
is $C^\infty$.
\item Four sample sizes are simulated $n\in\{200,300,500,1000\}$.
\item Three exponents are used $\gamma\in\{1,2,3\}$.
\end{itemize}
The following kernel is chosen 
$$
K(t)=\cos^2(\pi t /2)\mathbf{1}\{t\in[-1,1]\},
$$
with associated window width $h=4\hat\sigma(X) n^{-1/2}$ and with $p=n^{1/2}$.
The dependence of these sequences with respect to $n$ is chosen
according to Remark~\ref{rqun} with $\alpha=d=1$.
The multiplicative constant $4\hat\sigma(X)$ in $h$ is chosen heuristically.
The dependence with respect to the standard-deviation of $X$
is inspired from the density estimation case. The scale factor 4 was
chosen on the basis of intensive simulations.

\noindent Here, the experiment involves several steps:
\begin{itemize}
\item First, $m=500$ replications of the sample are simulated.  
\item For each of the $m$ previous set of points, the 
frontier estimator $\hat g_n$ is computed.  
\item The $m$ associated $L_1$ distances to $g$ are evaluated on a grid.  
\item The mean, smallest and largest $L_1$ errors are recorded.
\end{itemize}
\noindent Some results are depicted in~\cite{Nous},
where the best situation
(i.e. the estimation corresponding to the smallest $L_1$ error)
and the worst situation
(i.e. the estimation corresponding to the largest $L_1$ error)
are represented.
Note that, even in the worst situations, the empirical choices of sequences
$h$ and $p$ seem satisfying for all the considered frontiers
and densities of $X$. In fact, the worst situations are obtained when 
no points were simulated at the boundaries of the support.

\noindent Finally, the above estimator is compared to three other ones:
\begin{itemize}
\item The estimator $\hat g_n$ with $p=1$, which reduces to a rescaling
of the regression estimator, in a similar spirit as in~\cite{JacSuq2}.
\item Geffroy's estimator~\cite{Geffroy}, denoted by $\hat g_n^G$,
which is a step function based on the extreme values of the sample.
\item The kernel estimator $\hat g_n^K$ introduced in~\cite{ESAIM}, which is
a smoothed and bias-corrected version of Geffroy's estimator.
\end{itemize}
Results are summarized in Table~\ref{tab}. It appears that, when $\gamma$
increases, performances of all estimators decrease, since the simulated
points are getting more and more distant from the frontier function.
In the case $p=1$ and $\gamma=3$, one can see that $\hat g_n$ does
not converge to the true frontier when $n$ increases. This shows
that the condition $p\to\infty$ is necessary to obtain the convergence
of the estimator. Finally, note that in all the situations considered
in Table~\ref{tab}, $\hat g_n$ outperforms $\hat g_n^G$ and $\hat g_n^K$.

\section{Conclusion and further work}
\label{futur}

To conclude, let us note that, even though $\hat g_n$ converges
to the true frontier $g$ in case of non uniform conditional 
distributions, it is possible to design new estimators dedicated
to particular parametric models. For instance, in case
of model~(\ref{proba}), estimator $\hat g_n$ could be modified
to obtain
$$
\tilde g_{n,\gamma}(x)=\left(\frac{1}{\gamma B(1+p,\gamma)} \left.\sum_{i=1}^n K_h(x-X_i) Y_i^p \right/ 
\sum_{i=1}^n K_h(x-X_i) \right)^{1/p},
$$
where $B$ is the beta function defined by
$$
B(a,b) 	=	\int_0^1 u^{a-1}(1-u)^{b-1}du.
$$
Of course, $\hat g_n$ corresponds to the particular case $\tilde g_{n,1}$.
When $\gamma$ is assumed to be known, the new multiplicative constant
yields a very efficient bias correction, see~\cite{Nous}
for an illustration. A part of our future work will consist in
defining an estimator of $\gamma$ and plugging it into $\tilde g_{n,\gamma}$.
New asymptotic results will be established.
We also plan to investigate the asymptotic properties of 
local polynomial estimators based on the same ideas as those used for $\hat g_n$
and $\tilde g_{n,\gamma}$.

%-----------------------------------------------------------------------------
\section{Appendix: Auxiliary lemmas}
%--------------------------------------------------------------------------
\label{proofs}

The following lemma provides convenient bounds obtained by a specific
study of the functions $u\to |(1+u)^{p}-1| -2p|u|$ and 
$u\to(1+u)^{1/p}-1-\frac{1}{p}u$. The study is left to the
reader. Note that these bounds could not be directly derived from the Taylor
formulas $|(1+u)^{p}-1| =|pu+o(u)|$
and $\left|(1+u)^{1/p}-1-\frac{1}{p}u\right| =\left|\frac{1}
{2p}(\frac{1}{p}-1)u^{2}+o(u^{2})\right| $ where the dependence on $p$ of
$o(u)$ and $o(u^{2})$ is not precised.

\begin{Lem}
\label{lemtrois}
Suppose $p\geq 1$. 
\begin{description}
\item [(i)] Then, $p |u|\leq\ln2$ entails
$ |(1+u)^{p}-1| \leq2p| u|.  $
\item[(ii)] Let $C\geq2$. Then, $|u| <1/2$ entails
$\left| (1+u)^{1/p}-1-\frac{1}{p}u\right| \leq\frac{C}{p}u^{2}.$
\end{description}
\end{Lem}

%--------------------------------------------------------------------------
\noindent The next lemma is dedicated to the control of the local variations
of the frontier on a neighborhood of size $h$.  

\begin{Lem}
\label{lemquatre}
Suppose \Au, \Ad~hold. If $ph^\alpha\to0$ and $\|x-y\| \leq h$, then for sufficiently
large $n$,
$$
\left| \left(  \frac{g(x)}
{g(y)}\right)  ^{p}-1\right| \leq2\frac{L_{g}}{g_{\min}}ph^{\alpha},
$$
where $L_{g}$ is the Lipschitz constant of the function $g$. 
\end{Lem}
\proof
Take $u=\frac{g(x)}{g(y)}-1$ and observe that 
$p|u|\leq p\frac{L_{g}}{g_{\min}}\|x-y\|^{\alpha}$,
Thus, if $\|x-y\| \leq h$, and $ph^{\alpha}\rightarrow0$, we have 
$p|u|\leq\ln2$ for sufficiently large $n$. 
Then, following Lemma~\ref{lemtrois}(i), 
for sufficiently large $n$, we obtain
$$
\left| (1+u)^{p}-1\right| = \left| \left(  \frac{g(x)}
{g(y)}\right)  ^{p}-1\right| \leq2p|u| \leq
2\frac{L_{g}}{g_{\min}}ph^{\alpha},
$$
and the result is proved.  
\CQFD

%--------------------------------------------------------------------------
\noindent Lemma~\ref{lemcinq} is used to establish the
complete convergence of random variables ratio.  

\begin{Lem}
\label{lemcinq}
Let $S$, $T$ be real random variables,
$a$, $b$ non zero real numbers, and $0<\eta<1/2 $. Then,
$$
\left\{  \left\vert \frac{S}{T}-\frac{a}{b}\right\vert >4\eta\left\vert
\frac{a}{b}\right\vert \right\}  \subseteq\left\{  \left\vert \frac{S}%
{a}-1\right\vert >\eta\right\}  \cup\left\{  \left\vert \frac{T}%
{b}-1\right\vert >\eta\right\}  .
$$
\end{Lem}
\proof
Consider the following obvious equality:
\begin{equation}
\label{eqlemcinq}
\left(  \frac{S}{T}-\frac{a}{b}\right)  =\frac{a}{b}\left(  \frac{S}
{a}-1\right)  +\frac{a}{b}\left(  1-\frac{T}{b}\right)  +\left(  \frac{S}
{T}-\frac{a}{b}\right)  \left(  1-\frac{T}{b}\right) . 
\end{equation}
The triangular inequality yields for all $\eta>0$:
$$
\left\{  \left\vert \frac{S}{a}-1\right\vert \leq\eta\right\}  \cap\left\{
\left\vert \frac{T}{b}-1\right\vert \leq\eta\right\} 
\subseteq
\left\{
\left\vert \frac{S}{T}-\frac{a}{b}\right\vert \leq2\eta\left\vert \frac{a}
{b}\right\vert +\eta\left\vert \frac{S}{T}-\frac{a}{b}\right\vert \right\}  .
$$
Taking $0<\eta<1$, we obtain
$$
\left\{  \left\vert \frac{S}{a}-1\right\vert \leq\eta\right\}  \cap\left\{
\left\vert \frac{T}{b}-1\right\vert \leq\eta\right\} 
\subseteq
\left\{
\left\vert \frac{S}{T}-\frac{a}{b}\right\vert \leq\frac{2\eta}{1-\eta
}\left\vert \frac{a}{b}\right\vert \right\}.
$$
Finally, note that $\frac{2\eta}{1-\eta}<4\eta$ for $0<\eta<1/2$.
\CQFD

%--------------------------------------------------------------------------
\noindent The next three lemmas are of great use to deduce successively
the asymptotic normality of $\widehat g_n(x)$ from $\widehat r_n(x)$
and the asymptotic normality of $\widehat r_n(x)$ from $\widehat\varphi_n(x)$.

\begin{Lem}
\label{lemsix}
Let $x\in\R^d$.  
If ${\widehat{f}_{n}(x)}/{f(x)}\toP 1$ and 
${\widehat{\varphi}_{n}(x)}/{\varphi_{n}(x)}\toP 1$, then
$$
\left(  \frac{\widehat{r}_{n}(x)}{r_{n}(x)}-1\right)  =\left(  \frac{\widehat
{\varphi}_{n}(x)}{\varphi_{n}(x)}-1\right)  -\left(  \frac{\widehat{f}_{n}(x)}
{f(x)}-1\right)(1+o_{p}(1)).
$$
\end{Lem}
\proof The hypotheses yield $\frac{\widehat{r}_{n}(x)}{r_{n}(x)}
=\frac{\widehat{\varphi}_{n}(x)}{\widehat{f}_{n}(x)}/\frac{\varphi_{n}(x)}{f(x)}
\toP 1$. Thus it suffices to consider $S=\frac{\widehat
{\varphi}_{n}(x)}{\varphi_{n}(x)}$, $T=\frac{\widehat{f}_{n}(x)}{f(x)}$, and $a=b=1$ in the equality~(\ref{eqlemcinq}).
\CQFD
%--------------------------------------------------------------------------

\begin{Lem}
\label{lemsept}
Let $x\in\R^d$.  
If ${\widehat{f}_{n}(x)}/{f(x)}\toP 1$ and 
${\widehat{\varphi}_{n}(x)}/{\varphi_{n}(x)}\toP 1$, then
$$
\left(  \frac{\widehat{g}_{n}(x)}{g(x)}-1\right)  =\frac{1}{p}\left(
\frac{\widehat{r}_{n}(x)}{r_{n}(x)}-1\right)(1+o_{p}(1)).
$$
\end{Lem}
\proof
From the hypotheses, $w_{n}(x):=\frac{\widehat{r}_{n}(x)}{r_{n}(x)}-1=o_{p}(1)$.
Moreover, following Lemma~\ref{lemtrois}(ii), on the event
$\left\{  \left\vert w_{n}(x) \right\vert <1/2\right\}  $ we have:
$$
\Delta_{n}(x):=\left\vert \left(  \frac{\widehat{g}_{n}(x)}{g(x)}-1\right) 
-\frac {1}{p}\left(  \frac{\widehat{r}_{n}(x)}{r_{n}(x)}-1\right) \right\vert
=\left\vert \left(  1+w_{n}(x)\right)  ^{1/p}-1-\frac{w_{n}(x)}{p}\right\vert 
\leq C\frac{1}{p} w_{n}^{2}(x).
$$
We thus have, on the one hand,
$$
p\Delta_{n}(x)\mathbf{1}_{\left\{  \left\vert w_{n}(x)\right\vert
<1/2\right\}  }=o_{p}(w_{n}(x)).
$$
On the other hand, for all $\varepsilon>0$,
$$
\left\{ p\frac{\Delta_{n}(x)}{w_{n}(x)}\mathbf{1}_{\left\{  \left\vert
w_{n}(x)\right\vert \geq1/2\right\}  }> \varepsilon\right\}
\subseteq \left\{  \left\vert w_{n}(x)\right\vert \geq1/2\right\}
$$
leading to
$$
P\left\{  p\frac{\Delta_{n}(x)}{w_{n}(x)}\mathbf{1}_{\left\{  \left\vert
w_{n}(x)\right\vert \geq1/2\right\}  }> \varepsilon\right\}  \leq P\left\{  \left\vert
w_{n}(x)\right\vert \geq1/2\right\}  \rightarrow0,
$$
and thus
$$
p\frac{\Delta_{n}(x)}{w_{n}(x)}\mathbf{1}_{\left\{  \left\vert
w_{n}(x)\right\vert \geq1/2\right\}  }=o_{p}(w_{n}(x)) ,
$$
which completes the proof.
\CQFD

\begin{Lem}
\label{lemhuit}
Suppose that $nph^{d+2\alpha}\rightarrow0$ and $p/(nh^d)\rightarrow0$. 
Let us define
$$
\sigma_{n}^{-1}(x)=((2p+1)nh^d)^{1/2}
\left(\frac{f(x)}{\int_B K^{2}(t)dt}\right)^{1/2},
$$
and let $Q$ be an arbitrary distribution.  
Then, under \Au--\Ac, 
$$
\left\{
\frac{\sigma_{n}^{-1}(x)}{p}\left(\frac{\widehat{\varphi}_{n}(x)}{\varphi_{n}(x)
}-\frac{\E\widehat{\varphi}_{n}(x)}{\varphi_{n}(x)}\right) \tod Q
\right\}
\Longrightarrow
\left\{
\sigma_{n}^{-1}(x)\left(\frac{\widehat{g}_{n}(x)}{g(x)}-1\right) \tod Q
\right\}.
$$
\end{Lem}
\noindent
\proof
First, note that $nph^{d+2\alpha}\rightarrow0$ and $p/(nh^d)\rightarrow0$
imply $ph^\alpha\to0$.
Thus, from Corollary~\ref{coro}, $\widehat\varphi_n(x)/\varphi_n(x)\toP 1$.
Besides, $p/(nh^d)\to 0$ implies $nh^d\to\infty$, and thus, 
using a classical result on density estimation
(see for instance~\cite{Bosq}, Chapter~4, Theorem~II.1),
we have $\widehat f_n(x)/f(x)\toP 1$.
Lemma~\ref{lemsix} thus entails
\begin{eqnarray*}
\frac{\sigma_{n}^{-1}(x)}{p}\left(  \frac{\widehat{r}_{n}(x)}{r_{n}(x)}-1\right)
&=& \frac{\sigma_{n}^{-1}(x)}{p}\left(  \frac{\widehat
{\varphi}_{n}(x)}{\varphi_{n}(x)}-1\right)  - \frac{\sigma_{n}^{-1}(x)}{p}\left(  \frac{\widehat{f}_{n}(x)}
{f(x)}-1\right)(1+o_{p}(1)) \\
&=&\frac{\sigma_{n}^{-1}(x)}{p}\left(  \frac{\widehat{\varphi}_{n}(x)}{\varphi_{n}(x) }-\frac{\E\widehat{\varphi}_{n}(x)}{\varphi_{n}(x)}\right)  -\frac{\sigma_{n}^{-1}(x)}{p}\left(  \frac{\widehat{f}_{n}(x)}{f(x)}-\frac{\E\widehat{f}_{n}(x)}{f(x)}\right)(1+o_p(1))\\
\label{eqsept}
&+& \frac{\sigma_{n}^{-1}(x)}{p}\left(  \frac{\E \widehat{\varphi}_{n}(x)}{\varphi_{n}(x) }-1\right)  -\frac{\sigma_{n}^{-1}(x)}{p}\left(  \frac{\E\widehat{f}_{n}(x)}{f(x)}-1\right)(1+o_p(1)).
\end{eqnarray*}
Following Lemma~\ref{lemun}, we have,
$$
\frac{\sigma_{n}^{-1}(x)}{p}\left(\frac{\E\widehat{\varphi}_{n}(x)}{\varphi_{n}(x)}-1\right)  
=O\left(  \left({\frac{nh^d}{p}}\right)^{1/2}\right)  O\left(  ph^{\alpha}\right)
=O\left(\left({nph^{d+2\alpha}}\right)^{1/2}\right)=o(1) , 
$$
and from a classical result on density estimation
$
\E\widehat{f}_{n}(x)-f(x)=O(h^\alpha),
$
see~\cite{Collomb}, Proposition~2.1, 
we have
$$
\frac{\sigma_{n}^{-1}(x)}{p}\left(  \frac{\E\widehat{f}_{n}(x)}{f(x)}-1\right)
=O\left( \left({\frac{nh^d}{p}}\right)^{1/2}\right) O\left(  h^{\alpha}\right) 
=O\left( \left({np^{-1}h^{d+2\alpha}}\right)^{1/2}\right)=o(1).
$$
Consequently, 
$$
\frac{\sigma_{n}^{-1}(x)}{p}\left(  \frac{\widehat{r}_{n}(x)}{r_{n}(x)}-1\right)
= \frac{\sigma_{n}^{-1}(x)}{p}\left(  \frac{\widehat
{\varphi}_{n}(x)}{\varphi_{n}(x)}-1\right)  - \frac{\sigma_{n}^{-1}(x)}{p}\left(  \frac{\widehat{f}_{n}(x)}
{f(x)}-1\right)(1+o_{p}(1))+o_p(1).
$$
Again, using a classical result on density estimation,
$
\V(\widehat{f}_{n}(x))=O(1/(nh^d)),
$
see~\cite{Collomb}, Proposition~2.2, we have
$$
\V\left(  \frac{\sigma_{n}^{-1}(x)}{p} \frac{\widehat{f}_{n}(x)}{f(x)}\right)
=O\left({\frac{nh^d}{p}}\right) O\left(  \frac{1}{nh^d}\right)=O(1/p)=o(1),
$$
and thus
\begin{equation}
\label{eqquatre}
\frac{\sigma_{n}^{-1}(x)}{p}\left(  \frac{\widehat{r}_{n}(x)}{r_{n}(x)}-1\right)
= \frac{\sigma_{n}^{-1}(x)}{p}\left(  \frac{\widehat
{\varphi}_{n}(x)}{\varphi_{n}(x)}-1\right) +o_p(1).
\end{equation}
Suppose now that there exists a probability distribution $Q$
such that 
$$
\frac{\sigma_{n}^{-1}(x)}{p}\left(  \frac{\widehat{\varphi}_{n}(x)}
{\varphi_{n}(x)}-\frac{\E\widehat{\varphi}_{n}(x)}{\varphi_{n}(x)}\right)
\tod Q.
$$
From (\ref{eqquatre}), we deduce that 
$$
\frac{\sigma_{n}^{-1}(x)}{p}\left( 
\frac{\widehat{r}_{n}(x)}{r_{n}(x)}-1\right)  \tod Q.
$$
Finally, from Lemma~\ref{lemsept} we can conclude that
$$
\sigma_{n}^{-1}(x)\left(
\frac{\widehat{g}_{n}(x)}{g(x)}-1\right)  \tod Q,
$$
and the result is proved.
\CQFD

\bibliographystyle{plain}

\begin{thebibliography}{10}

\bibitem{Aragon} Y.~Aragon, A.~Daouia and C.~Thomas-Agnan.
Nonparametric frontier estimation: a conditional quantile-based approach. 
{\em Journal of Econometric Theory}, 21(2):358--389, 2005.

\bibitem{Bosq}
D.~Bosq and J.P. Lecoutre.
\newblock {\em Th\'eorie de l'estimation fonctionnelle}.
\newblock Economie et Statistiques avanc\'ees. Economica, Paris, 1987.

\bibitem{Cazals}
C.~Cazals, J.-P.~Florens and L.~Simar.
Nonparametric frontier estimation: A robust approach. 
{\em Journal of Econometrics}, 106(1):1--25, 2002.

\bibitem{Collomb}
G.~Collomb.
\newblock {\em Estimation non param\'etrique de la r\'egression par la
  m\'ethode du noyau}.
\newblock PhD thesis, Universit\'e Paul Sabatier de Toulouse, 1976.

\bibitem{DST}
D.~Deprins, L.~Simar, and H.~Tulkens.
\newblock Measuring labor efficiency in post offices.
\newblock In P.~Pestieau M.~Marchand and H.~Tulkens, editors, {\em The
  Performance of Public Enterprises: Concepts and Measurements}. North Holland
  ed, Amsterdam, 1984.


\bibitem{EMBR} P.~Embrechts, C.~Kl\"uppelberg, and T.~Mikosch.  {\em Modelling extremal events}, Springer, 1997.

\bibitem{Farrel}
M.J. Farrel.
\newblock The measurement of productive efficiency.
\newblock {\em Journal of the Royal Statistical Society A}, 120:253--281, 1957.

\bibitem{LivreToulouse}
F.~Ferraty and P.~Vieu.
\newblock {\em Nonparametric modelling for functional data}.
\newblock Springer, 2005.

\bibitem{ISUPLaurent}
L.~Gardes.
\newblock Estimating the support of a {P}oisson process via the
  {F}aber-{S}hauder basis and extreme values.
\newblock {\em Publications de l'Institut de Statistique de l'Universit\'{e} de
  Paris}, XXXXVI:43--72, 2002.

\bibitem{Geffroy}
J.~Geffroy.
\newblock Sur un probl\`{e}me d'estimation g\'{e}om\'{e}trique.
\newblock {\em Publications de l'Institut de Statistique de l'Universit\'{e} de
  Paris}, XIII:191--210, 1964.

\bibitem{Gijbels2}
I.~Gijbels, E.~Mammen, B.~U. Park, and L.~Simar.
\newblock {On estimation of monotone and concave frontier functions.}
\newblock {\em Journal of the American Statistical Association},
  94(445):220--228, 1999.

\bibitem{Russe2}
S.~Girard, A.~Iouditski, and A.~Nazin.
\newblock ${L}_1$-optimal nonparametric frontier estimation via linear
  programming.
\newblock {\em Automation and Remote Control}, 66(12):2000--2018, 2005.

\bibitem{Scandi}
S.~Girard and P.~Jacob.
\newblock Extreme values and {H}aar series estimates of point process
  boundaries.
\newblock {\em Scandinavian Journal of Statistics}, 30(2):369--384, 2003.

\bibitem{JSPI}
S.~Girard and P.~Jacob.
\newblock Projection estimates of point processes boundaries.
\newblock {\em Journal of Statistical Planning and Inference}, 116(1):1--15,
  2003.

\bibitem{ESAIM}
S.~Girard and P.~Jacob.
\newblock Extreme values and kernel estimates of point processes boundaries.
\newblock {\em ESAIM: Probability and Statistics}, 8:150--168, 2004.


\bibitem{Nous}
S.~Girard and P.~Jacob.
\newblock Frontier estimation via kernel regression on high power-transformed data. 
\newblock {\em Journal of Multivariate Analysis}, 99:403--420, 2008.

\bibitem{JSPI2}
S.~Girard and L.~Menneteau.
\newblock Central limit theorems for smoothed extreme value estimates of point
  processes boundaries.
\newblock {\em Journal of Statistical Planning and Inference}, 135(2):433--460,
  2005.

\bibitem{Hall2}
P.~Hall, M.~Nussbaum, and S.~Stern.
\newblock {On the estimation of a support curve of indeterminate sharpness.}
\newblock {\em Journal of Multivariate Analysis}, 62(2):204--232, 1997.

\bibitem{Hall3}
P.~Hall and B.~U. Park.
\newblock {Bandwidth choice for local polynomial estimation of smooth
  boundaries.}
\newblock {\em Journal of Multivariate Analysis}, 91(2):240--261, 2004.

\bibitem{Hall}
P.~Hall, B.~U. Park, and S.~E. Stern.
\newblock {On polynomial estimators of frontiers and boundaries.}
\newblock {\em Journal of Multivariate Analysis}, 66(1):71--98, 1998.

\bibitem{Har}
W.~H\"ardle, B.~U. Park, and A.~B. Tsybakov.
\newblock Estimation of a non sharp support boundaries.
\newblock {\em Journal of Multivariate Analysis}, 43:205--218, 1995.

\bibitem{JacSuq}
P.~Jacob and P.~Suquet.
\newblock Estimating the edge of a {P}oisson process by orthogonal series.
\newblock {\em Journal of Statistical Planning and Inference}, 46:215--234,
  1995.

\bibitem{JacSuq2}
P.~Jacob and P.~Suquet.
\newblock Regression and edge estimation.
\newblock {\em Statistic and Probability Letters}, 27:11--15, 1996.

\bibitem{Keith}
K.~Knight.
\newblock {Limiting distributions of linear programming estimators.}
\newblock {\em Extremes}, 4(2):87--103, 2001.

\bibitem{KorTsy3}
A.~Korostelev, L.~Simar, and A.~B. Tsybakov.
\newblock Efficient estimation of monotone boundaries.
\newblock {\em The Annals of Statistics}, 23:476--489, 1995.

\bibitem{KorTsy}
A.P. Korostelev and A.B. Tsybakov.
\newblock {\em Minimax theory of image reconstruction}, volume~82 of {\em
  Lecture Notes in Statistics}.
\newblock Springer-Verlag, New-York, 1993.

\bibitem{MamTsy}
E.~Mammen and A.~B. Tsybakov.
\newblock Asymptotical minimax recovery of set with smooth boundaries.
\newblock {\em The Annals of Statistics}, 23(2):502--524, 1995.

\end{thebibliography}

%-----------------------------------------------------------------------------
\begin{table}
\begin{tabular}{|c|c|c|c|c|}
\hline
\multicolumn{5}{|c|}{$\gamma=1$}\\
\hline
$n$ & $\hat g_n$ with $p\to\infty$ & $\hat g_n$ with $p=1$ & $\hat g_n^{K}$ & $\hat g_n^{G}$ \\
\hline
200 & 0.121 {\it [0.051, 0.237]} & 0.651 {\it [0.407, 0.907]} & 0.134 {\it [0.056, 0.261]} & 0.183 {\it [0.080, 0.334]}\\
\hline
300 & 0.100 {\it [0.049, 0.184]} & 0.636 {\it [0.445, 0.831]} & 0.111 {\it [0.061, 0.219]} & 0.157 {\it [0.073, 0.300]}\\
\hline
500 & 0.078 {\it [0.042, 0.138]} & 0.627 {\it [0.441, 0.813]} & 0.087 {\it [0.046, 0.168]} & 0.128 {\it [0.064, 0.234]}\\
\hline
1000& 0.057 {\it [0.028, 0.112]} & 0.616 {\it [0.486, 0.752]} & 0.062 {\it [0.033, 0.117]} & 0.093 {\it [0.049, 0.158]}\\
\hline
\end{tabular}

\begin{tabular}{|c|c|c|c|c|}
\hline
\multicolumn{5}{|c|}{$\gamma=2$}\\
\hline
$n$ & $\hat g_n$ with $p\to\infty$ & $\hat g_n$ with $p=1$ & $\hat g_n^{K}$ & $\hat g_n^{G}$ \\
\hline
200 & 0.321 {\it [0.197, 0.496]} & 0.575 {\it [0.415, 0.759]} & 0.337 {\it [0.180, 0.519]} & 0.426 {\it [0.269, 0.591]}\\
\hline
300 & 0.297 {\it [0.194, 0.457]} & 0.562 {\it [0.399, 0.755]} & 0.311 {\it [0.171, 0.490]} & 0.393 {\it [0.255, 0.569]}\\
\hline
500 & 0.262 {\it [0.169, 0.379]} & 0.545 {\it [0.429, 0.667]} & 0.275 {\it [0.172, 0.380]} & 0.347 {\it [0.251, 0.452]}\\
\hline
1000& 0.226 {\it [0.153, 0.303]} & 0.533 {\it [0.463, 0.623]} & 0.240 {\it [0.152, 0.336]} & 0.293 {\it [0.200, 0.388]}\\
\hline
\end{tabular}

\begin{tabular}{|c|c|c|c|c|}
\hline
\multicolumn{5}{|c|}{$\gamma=3$}\\
\hline
$n$ & $\hat g_n$ with $p\to\infty$ & $\hat g_n$ with $p=1$ & $\hat g_n^{K}$ & $\hat g_n^{G}$ \\
\hline
200 & 0.526 {\it [0.331, 0.709]} & 0.749 {\it [0.627, 0.888]} & 0.550 {\it [0.340, 0.724]} & 0.624 {\it [0.410, 0.780]}\\
\hline
300 & 0.496 {\it [0.363, 0.669]} & 0.744 {\it [0.632, 0.865]} & 0.523 {\it [0.371, 0.687]} & 0.591 {\it [0.452, 0.739]}\\
\hline
500 & 0.457 {\it [0.366, 0.590]} & 0.741 {\it [0.649, 0.817]} & 0.486 {\it [0.375, 0.620]} & 0.545 {\it [0.434, 0.668]}\\
\hline
1000& 0.410 {\it [0.315, 0.505]} & 0.742 {\it [0.685, 0.817]} & 0.442 {\it [0.327, 0.531]} & 0.486 {\it [0.375, 0.573]}\\
\hline
\end{tabular}
\caption{Comparison between $L_1$ errors obtained with four different estimators. The mean error is indicated as well as the range between the minimum and the maximum error.
The experiments are conducted on a $B(2,2)$ covariate, with frontier function $g_2$.} 
\label{tab}
\end{table}

\end{document}